# Theoretical Study of Phosphorene Tunneling Field Effect Transistors

Jiwon Chang and Chris Hobbs

SEMATECH, 257 Fuller Rd #2200, Albany, NY 12203, USA

In this work, device performances of tunneling field effect transistors (TFETs) based on phosphorene are explored via self-consistent atomistic quantum transport simulations. Phosphorene is an ultra-thin two-dimensional (2-D) material with a direct band gap suitable for TFETs applications. Our simulation shows that phosphorene TFETs exhibit subthreshold slope ($SS$) below 60 mV/dec and a wide range of on-current depending on the transport direction due to highly anisotropic band structures of phosphorene. By benchmarking with monolayer $MoTe_2$ TFETs, we predict that phosphorene TFETs oriented in the small effective mass direction can yield much larger on-current at the same on-current/off-current ratio than monolayer $MoTe_2$ TFETs. It is also observed that a gate underlap structure is required for scaling down phosphorene TFETs in the small effective mass direction to suppress the source-to-drain direct tunneling leakage current.

Tunneling field effect transistors (TFETs) have gained a lot of attention over the past few years due to the ability to achieve subthreshold slope (*SS*) steeper than 60 mV/dec, a fundamental limit in metal-oxide-semiconductor field effect transistors (MOSFETs) [1]. With the reduced *SS*, TFETs provide a promising path to scale down the supply voltage, and hence reduce static and dynamic power consumptions. There have been numerous researches to explore band-to-band tunneling principles in silicon [2], germanium [3], and III-V materials [4,5] to obtain *SS* less than 60 mv/dec with the high on-current/off-current ratio. In recent years, atomically thin two-dimensional (2-D) layered materials such as graphene, topological insulators (TIs), and transition metal dichalcogenides (TMDs) have found great significance due to the outstanding electrostatic integrity. With the improved gate control over the channel, TFETs based on 2-D materials have become promising options for ultra-low power devices. Graphene nanoribbon (GNR) TFETs with a tunable band gap have been simulated to exhibit excellent device performances [6]. However, it is challenging to control the width of GNR perfectly to avoid the line edge roughness issue [7,8]. TFETs using TMDs and TIs whose band gap sizes are controlled by the thickness would be analogous to GNR TFETs, but without the sensitivity to ribbon width and edge roughness. The potential of ultra-thin TIs $Bi_2Se_3$ as a channel material in TFETs has been explored [9,10]. A theoretical study has proven the possibility of monolayer TMDs for TFETs, but with limited device performances [11]. Very recently, another 2-D layered material, phosphorene, was obtained experimentally and proposed for FETs applications [12,13]. Phosphorene, a single layer of black phosphorus arranged in a puckered honeycomb structure, has a direct and thickness dependent band gap [12,14,15,16,17]. Unlike other 2-D materials, phosphorene exhibits a high degree of anisotropy in the band structures [12,14,15,17,18]. This unique anisotropy is helpful to enhance the device performance by providing a small transport mass leading to a high carrier velocity as well as a large transverse mass leading to a large density of states. In this work, we evaluated device performances of TFETs based on phosphorene through ballistic quantum transport simulations. Due to the anisotropic band structure, device performances of phosphorene TFETs were found to be sensitive to the phosphorene crystal orientation. We also simulated device performances of monolayer $MoTe_2$ TFETs under the same biasing conditions and revealed that phosphorene TFETs can outperform. Finally, we report the scaling behavior of phosphorene TFETs.

The primitive unit cell of phosphorene is rectangular with a four-atom basis, as shown by the top view of phosphorene in Fig. 1(a). We investigated band structures through density functional theory (DFT) calculations with the OPENMX code [19] using the linear combination of pseudoatomic orbital (PAO) method. PAO basis set for the phosphorus atom was carefully chosen to reproduce the band structure obtained by plane wave based DFT calculations [12,14,17,18]. We adopted the local density approximation (LDA) [20] for exchange-correlation energy functional and used a kinetic energy cutoff of 300 Ryd and 7×7×1 $k$-mesh. Since it has been reported that DFT optimized lattice parameters are very close to the experimentally measured values [12,14,16,17,18], we employed experimental bulk lattice parameters of 4.38 and 3.31 Å for the in-plane lattice constants $a_x$ and $a_y$ (Fig. 1(a)) to construct the phosphorene structures [21]. Similar to the previous studies, we found a direct band gap at the Γ point and the calculated band gap size is 0.93 eV close to other theoretical reports [12,17,18]. It is well known that DFT shows the significant dependence of band gap size on the energy functionals. Indeed, band gap values with PBE functional are 0.84 [14] and 0.91 [17,18] eV while using HSE06 functional results in 1.0 [12], 1.52 [14] and ~1.5 eV [17]. Experimental observations also exhibit certain variations such as 1.31 [22] and 1.45 [12] eV via photoluminescence spectra and 1.0 eV from the transfer characteristics of phosphorene FETs [16]. Since our simulation produces the band gap quite close to the experimental value, obtained through the transfer characteristics of phosphorene FETs, our calculated band structures are valid for further device simulations. Band structures are highly anisotropic as predicted in other reports [12,14,15,17,18]. From Fig. 1(c), effective masses of electron and hole are very small ($m_e^* = 0.115 \times m_e$ and $m_h^* = 0.12 \times m_e$) in $x$-direction while those increase drastically ($m_e^* = 1.17 \times m_e$ and $m_h^* = 1.65 \times m_e$) in $y$-direction. Tight-binding (TB) potentials used in the quantum transport simulation were obtained from DFT using maximally localized Wannier functions (MLWFs) [23]. The Wannier functions and hopping potentials were calculated directly from the DFT Kohn-Sham orbitals and potential also using OPENMX. As in Fig. 1(c), the resulting TB band structures match well with the original DFT band structures.

For the transport calculation, we defined a series of rectangular unit cells in the simulation region along the transport direction. Two transport directions $x$ (small effective mass) and $y$ (large effective mass) shown in the top view of phosphorene in Fig. 1(a) were considered. A double gate (DG) structure with the gate dielectric of 2.5 nm $HfO_2$ ($\kappa = 25$) was simulated as illustrated in Fig. 1(b). We employed a p-i-n (p-type, intrinsic, n-type) doping profile for source-channel-drain. Dielectric constant of 10 was adopted for phosphorene [24]. We assumed no work function difference between the gate and the channel for simplicity. To model ballistic quantum transport, propagating wave functions were obtained using a recursive scattering matrix approach [25]. Source and drain leads are in the thermal equilibrium with the externally applied bias, and are used for injecting eigenmodes. Periodic boundary conditions were imposed in the device width direction. The total current was calculated by integrating the transmission probabilities over energy with the Fermi function weight. Ballistic transport equation was solved iteratively together with the Poisson's equation until self-consistency between charge density and electrostatic potential is achieved. All simulations were performed at 300 K.

We first investigate the device performances of phosphorene TFETs in the small effective mass direction ($x$-direction). Fig. 3(a) shows the transfer characteristics of 20 nm gate length device without gate underlap ($L_{UL}$ = 0 nm) for different source doping concentrations at $V_{DS}$ = 0.5 V. We keep the drain doping concentration $N_D$ at $1\times10^{13}$/cm$^2$ while increasing the source doping concentration $N_S$ from $1\times10^{13}$/cm$^2$ up to $5\times10^{13}$/cm$^2$. We define $V_{GS}$ so that the off-current $I_{off}$ is $10^{-5}$ µA/µm at $V_{GS}$ = 0 V with $N_S$ = $5\times10^{13}$/cm$^2$ to satisfy the ITRS (International Technology Roadmap for Semiconductors) requirement for low standby power applications [26]. In Fig. 3(a), all cases exhibit *SS* less than 60 mV/dec, the theoretical limit for conventional MOSFETs at 300K. The current at high $V_{GS}$ is enhanced by more than 200 times by increasing the doping concentration in the source from $1\times10^{13}$/cm$^2$ up to $5\times10^{13}$/cm$^2$ since a higher doping concentration reduces the screening length for tunneling. With $N_S$ = $5\times10^{13}$/cm$^2$, $SS \approx 28$ mV/dec, $I_{on}$ = 47 µA/µm at $V_{GS}$ = 0.5 V, $I_{off}$ = $10^{-5}$ µA/µm at $V_{GS}$ = 0.0 V, and $I_{on}/I_{off}$ ratio more than $10^6$ with a supply voltage $V_{DD}$ = 0.5 V are achieved. We may expect to lower the minimum leakage current with a higher level of source doping which suppresses the thermionic current by reducing the number of minority carriers above the conduction band (CB) edge in the source, but the minimum leakage current also increases with the higher source doping concentration shown in Fig. 2(a). To understand the components of minimum current, CB and valence band (VB) edges profiles along the device for $N_S$ = $5\times10^{13}$, $4\times10^{13}$, $3\times10^{13}$ and $2\times10^{13}$/cm$^2$ at $V_{GSmin}$ (the gate voltage at the minimum current) indicated in Fig. 2(a) are plotted in Fig. 2(b). Corresponding energy resolved current densities are shown in Fig. 2(c) which reveals three components of leakage current: electron thermionic current from the source CB edge, hole thermionic current from the drain VB edge and source-to-drain tunneling current. From Fig. 2(c), the electron thermionic current above the source CB edge decreases with the heavier source doping since CB edge moves away from the source Fermi level as seen in Fig. 2(b). On the other hand, the hole thermionic current below the drain VB edge remains unchanged due to the fixed drain doping concentration. However, the last component of leakage current, the source-to-drain tunneling current, increases for the higher source doping level since CB edge becomes closer to VB edge in the source-channel interface region in Fig. 2(b), leading to the enhanced tunneling efficiency. In this large band gap material aligned in the small effective mass direction, the leakage current is dominated by the source-to-drain tunneling current rather than the thermionic current as shown in Fig. 2(c). Therefore, even though a higher source doping concentration yields lower thermionic current, it results in the increase of overall minimum leakage current.

Device performances of phosphorene TFETs in the large effective mass direction ($y$-direction) are simulated and compared with those in the small effective mass direction ($x$-direction). Fig. 3(a) compares the transfer characteristics of 20 nm gate length device without gate underlap ($L_{UL}$ = 0 nm) in the transport direction $x$ (black circle) and $y$ (purple star) at $V_{DS}$ = 0.5 V. Corresponding plots of $I_{on}$ as a function of $I_{on}/I_{off}$ ratio and $I_{off}$ at the power supply voltage $V_{DD}$ = 0.5 V ($V_{on}$–$V_{off}$ = $V_{DD}$) are shown in Fig. 4(a) and 4(b), respectively. Source and drain doping concentrations of $5\times10^{13}$/cm$^2$ and $1\times10^{13}$/cm$^2$, respectively, are used since high $I_{on}/I_{off}$ ratio and good *SS* are obtained in the transport direction $x$ as discussed above. From Fig. 3(a), current is significant suppressed in $y$-direction as the large effective mass translates to the low tunneling probability. As a result, phosphorene TFETs can provide more than $10^8$ times higher $I_{on}$ in $x$-direction than in $y$-direction at the same $I_{on}/I_{off}$ ratio as seen in Fig. 4(a). This comparison clearly highlights the need to identify the crystal orientation of phosphorene for optimal device performances. To benchmark phosphorene TFETs, we carry out quantum transport simulations of monolayer $MoTe_2$ TFETs using the same simulation approach and device parameters of phosphorene TFETs. TB parameters for monolayer $MoTe_2$ used in our previous work [27] is adopted. Among several Mo based TMDs, $MoTe_2$ is chosen since it exhibit the smallest band gap (~1.1 eV). Transport only in the direction from $\Gamma$ to K in Fig. 1(d) is considered, but we expect that nearly isotropic carrier effective masses ($m_e^* \sim 0.64 \times m_e$ and $m_h^* \sim 0.78 \times m_e$) of monolayer $MoTe_2$ result in similar transfer characteristics irrespective of transport directions. Simulation results of monolayer $MoTe_2$ TFETs are compared with those of phosphorene TFETs in Fig. 3(a) and Fig. 4, suggesting that phosphorene TFETs in the small effective mass direction outperforms monolayer $MoTe_2$ TFETs. Phosphorene TFETs exhibit approximately $10^3$ times larger $I_{on}$ at the same $I_{on}/I_{off}$ ratio compared with monolayer $MoTe_2$ TFETs when phosphorene is aligned in the small effective mass direction.

The scaling limit of phosphorene TFETs in the transport direction $x$ is explored by simulating 15 and 10 nm gate length devices with the same parameters of 20 nm one. From the transfer characteristics of 20 (black circle), 15 (black triangle) and 10 (black square) nm gate length devices in Fig. 3(a), as the gate length $L_G$ is scaled below 20 nm, the minimum leakage current increases exponentially, and reaches more than $10^{-5}$ μA/μm as $L_G$ shrinks to 10 nm. As a result, the maximum achievable $I_{on}/I_{off}$ ratio drops down to about $5 \times 10^5$ with $L_G$ = 10 nm as observed in Fig. 4(a). To explain the drastic increase of minimum leakage current for $L_G$ < 20 nm, we show CB and VB edges profiles (Fig. 3(b)) along with the energy resolved current densities (Fig. 3 (c)) for 20 (blue circle), 15 (green triangle) and 10 (red square) nm gate length devices, respectively. CB and VB edges profiles in Fig. 3(b) are at $V_{GSmin}$, the gate voltage at the minimum current in the 20 nm device, indicated in Fig. 3(a). Since the device parameters are the same but only with different gate lengths, energies of CB and VB edges in the source, channel and drain, respectively, are almost same for all three gate lengths. Therefore, electron and hole thermionic currents above the source CB edge and below the drain VB edge, respectively, are independent of the gate length as in Fig. 3(c). However, as the gate length is scaled down, the width of tunneling between VB edge in the source-channel interface and CB edge in the channel-drain interface is reduced, narrowing down the tunneling barrier for the source-to-drain direct tunneling leakage current. Due to very small electron and hole effective masses in the $x$-direction transport, the direct tunneling leakage current rapidly increases with $L_G < 20$ nm in Fig. 3(c), thereby limiting the gate length scalability. In order to suppress the minimum current in the 10 nm gate length device, we use a gate underlap of 10 nm at the drain end of the channel. Introducing gate underlap results in a nearly linear potential drop in the ungated region of the channel, which increases the width of source-to-drain direct tunneling barrier near the drain Fermi level, as shown by band edges profiles (diamond brown) in Fig. 3(b). The increase of tunneling barrier thickness leads to an exponential decrease of the source-to-drain tunneling leakage current, as seen by the current spectrum (diamond brown) in Fig. 3(c), hence improving the maximum achievable $I_{on}/I_{off}$ ratio up to $10^7$ by pushing the minimum current down to about $10^{-6}$ µA/µm as shown in Fig. 4. As a result, $I_{off}$ requirement of $10^{-5}$ µA/µm by ITRS for the low standby power applications can be fulfilled in the 10 nm gate length device with $I_{on}/I_{off}$ ratio > $10^6$ as indicated by the red dash line in Fig. 4.

    We note that these simulations are within the approximation of ballistic transport. Including scattering processes, especially electron-phonon scattering, may impose limits on the achievable minimum leakage current and SS. Electron-phonon scattering could affect the

simulation results of ballistic transport in the long channel device or in the large effective mass direction transport where the majority of leakage current is attributed to the phonon-assisted band-to-band tunneling with the negligible direct source-to-drain tunneling [28]. In the short channel limit or in the small effective mass direction transport, however, the effect of phonon assisted tunneling is limited since the source-to-drain tunneling is dominant in the leakage current as in Fig. 3(c).

In conclusion, we have assessed device performances of phosphorene TFETs through the self-consistent ballistic quantum transport simulation. Due to the anisotropic band structures of phosphorene, device performances substantially depend on the transport direction. By changing the transport direction from the large to small effective mass, $I_{on}$ can be boosted more than eight orders of magnitude due to the enhanced tunneling efficiency. Compared with monolayer $MoTe_2$ TFETs, phosphorene TFETs aligned in the small effective mass direction achieves three orders of magnitude larger $I_{on}$ at the same $I_{on}/I_{off}$ ratio, suggesting superiority of phosphorene to monolayer $MoTe_2$ for TFETs application. In the scaling limit, a properly designed gate underlap is needed as the device scalability in the small effective mass direction suffers from the source-to-drain direct tunneling current.

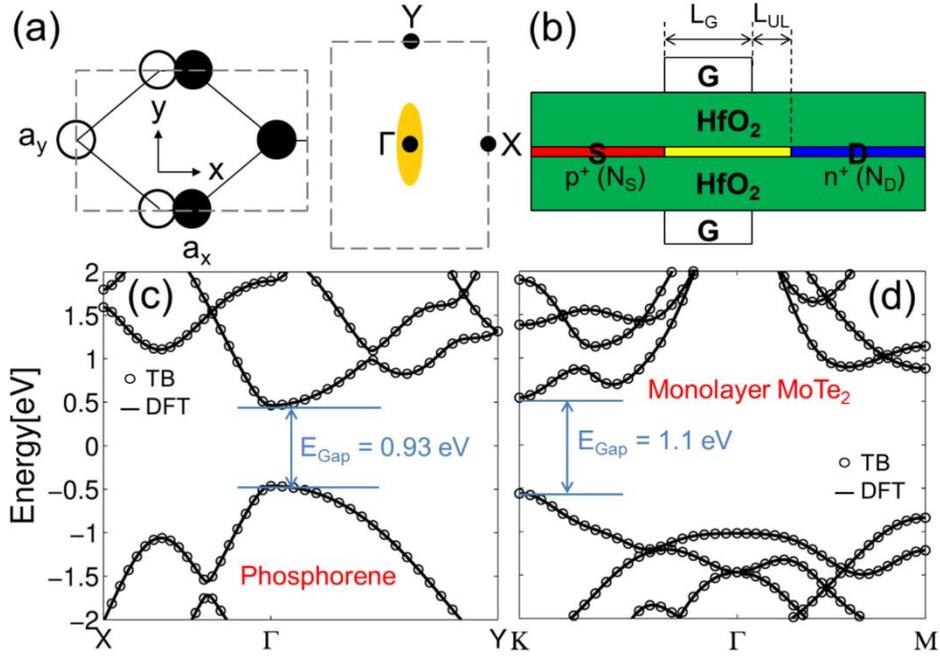

**FIG. 1**. (a) Top view of phosphorene showing a rectangular unit cell and corresponding 1$^{st}$ Brillouin Zone (BZ) with high symmetric points. (b) Simulated device structure of p-i-n phosphorene DG TFETs. $L_G$ and $L_{UL}$ are the gate length and the gate underlap length, respectively. Doping concentrations of source and drain are represented by $N_S$ and $N_D$, respectively. Band structures of (c) phosphorene and (d) monolayer MoTe$_2$ along the high symmetric paths in the rectangular BZ and hexagonal BZ, respectively, calculated from DFT (solid lines) and from TB Hamiltonian (circles).

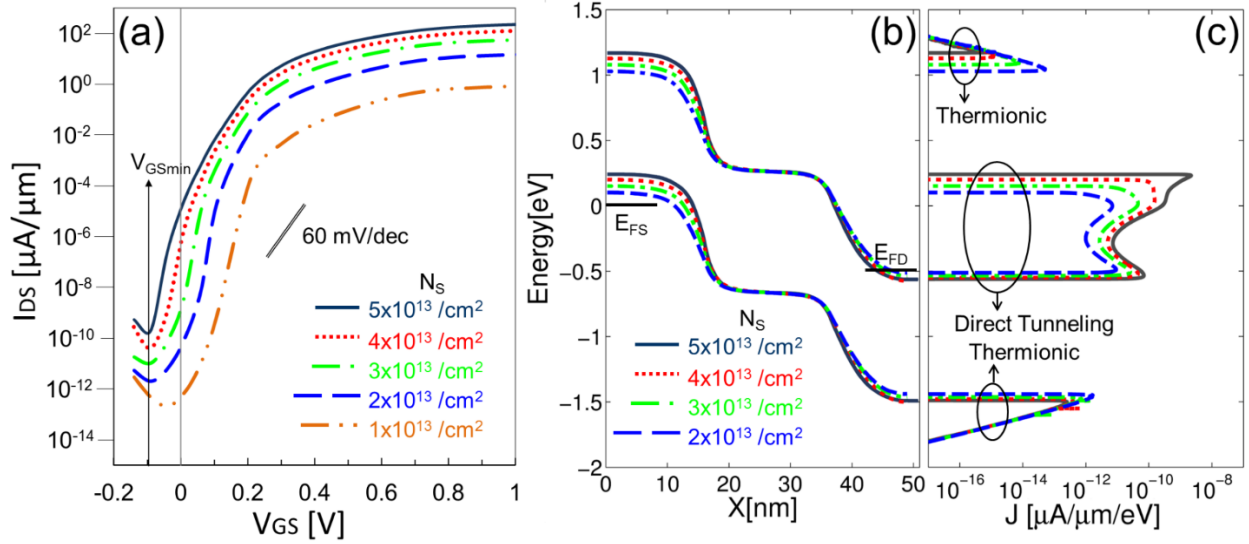

**FIG. 2**. (a) Transfer characteristics of phosphorene TFETs in the transport direction $x$ for different source doping concentrations $N_S$ = 5×10$^{13}$, 4×10$^{13}$, 3×10$^{13}$, 2×10$^{13}$ and 1×10$^{13}$/cm$^2$ at $V_{DS}$ = 0.5 V with $L_G$ = 20 nm and $L_{UL}$ = 0 nm. Drain doping concentration $N_D$ is kept at 1×10$^{13}$/cm$^2$. (b) CB and VB edges profiles along the device for $N_S$ = 5×10$^{13}$, 4×10$^{13}$, 3×10$^{13}$ and 2×10$^{13}$/cm$^2$ at $V_{GSmin}$ in (a). $E_{FS}$ and $E_{FD}$ represent Fermi levels in the source and drain, respectively. (c) Corresponding energy resolved current densities. The electron thermionic current from the source CB edge decreases for the higher $N_S$ while the source-to-drain direct tunneling current increases.

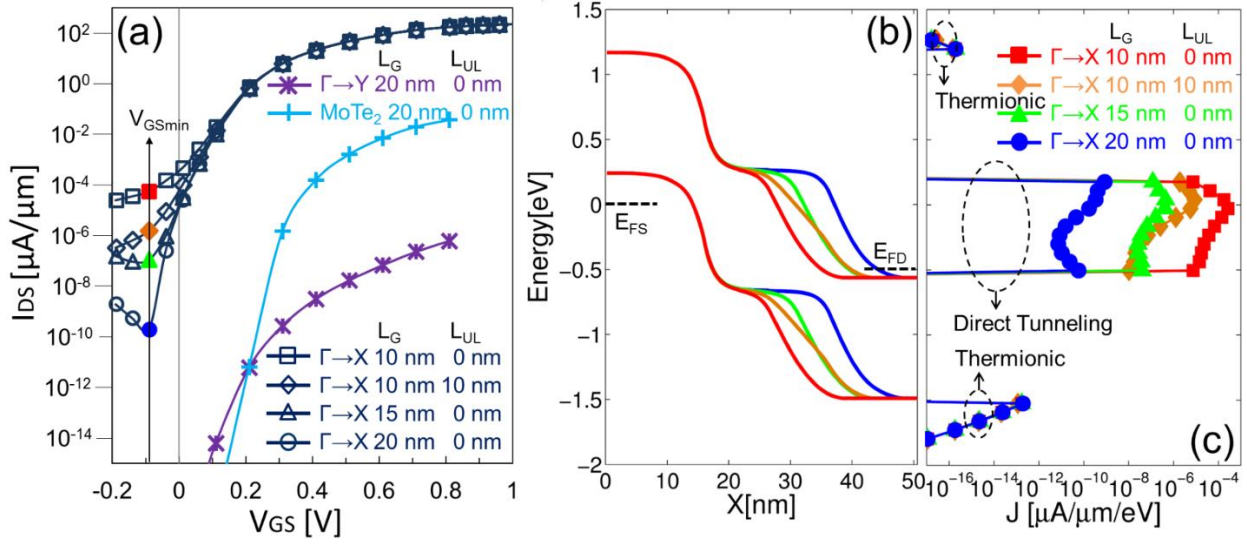

**FIG. 3**. (a) Transfer characteristics of phosphorene TFETs in the transport direction $x$ for different channel lengths $L_G$ = 10, 15 and 20 nm with gate underlap $L_{UL}$ = 0 or 10 nm at $V_{DS}$ = 0.5 V. Simulation results of phosphorene TFETs in the transport direction $y$ and monolayer MoTe$_2$ TFETs with $L_G$ = 20 nm and $L_{UL}$ = 0 nm are shown for comparison. Source and drain doping concentrations are $5\times10^{13}$/cm$^2$ and $1\times10^{13}$/cm$^2$, respectively, for all cases. (b) CB and VB edges profiles along the device for $L_G$ = 10, 15 and 20 nm with $L_{UL}$ = 0 or 10 nm at $V_{GSmin}$ in (a). $E_{FS}$ and $E_{FD}$ represent Fermi levels in the source and drain, respectively. (c) Corresponding energy resolved current densities. The electron and hole thermionic current from the source CB edge and drain VB edge, respectively, remain same while the source-to-drain direct tunneling current increases for shorter gate length.

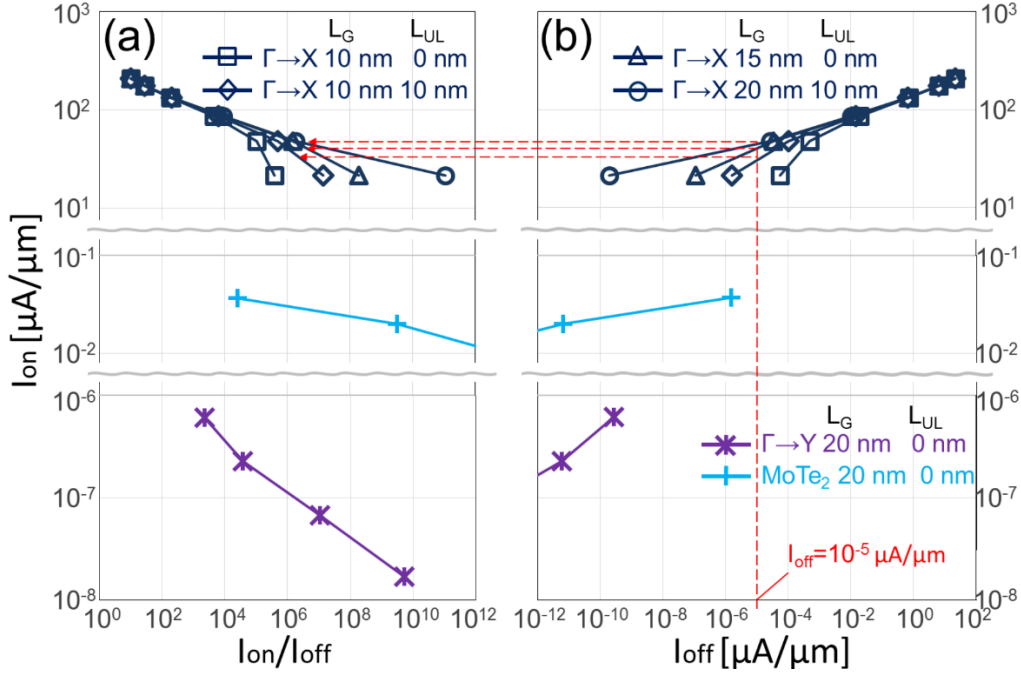

**FIG. 4**. (a) $I_{on}$ vs. $I_{on}/I_{off}$ ratio and (b) $I_{on}$ vs. $I_{off}$ at $V_{on} - V_{off} = V_{DD} = 0.5$ V for phosphorene TFETs in the transport direction $x$ for different channel lengths $L_G$ = 10, 15 and 20 nm with gate underlap $L_{UL}$ = 0 or 10 nm. Simulation results of phosphorene TFETs in the transport direction $y$ and monolayer MoTe$_2$ TFETs with $L_G$ = 20 nm and $L_{UL}$ = 0 nm are shown for comparison. $I_{on}$ at $I_{off} = 10^{-5}$ µA/µm (ITRS requirement for low standby power applications) and corresponding $I_{on}/I_{off}$ ratio are indicated in (b) and (a), respectively.

REFERENCES

1. A. C. Seabaugh and Q. Zhang, Proc. IEEE 98, 2095 (2010).

2. W. Y. Choi, B.-G. Park, J. D. Lee, and T.-J. King Liu, Electron Dev. Lett. 28, 743 (2007).

3. T. Krishnamohan, D. Kim, S. Raghunathan, and K. Saraswat, IEDM Tech. Dig., 947 (2008).

4. U. E. Avci, R. Rios, K. Kuhn, and I. Young, Proc. Symp. VLSI Tech., 124, (2011).

5.	J. Knoch and J. Appenzeller, IEEE Electron Device Lett. 31, 305 (2010).

6.	P. Zhao, J. Chauhan, and J. Guo, Nano Lett. 9, 684 (2009).

7.	D. Basu, M. J. Gilbert, L. F. Register, S. K. Banerjee, and A. H. MacDonald, Appl. Phys. Lett. 92, 042114 (2008).

8.	M. Luisier and G. Kimeck, Appl. Phys. Lett. 94, 223505 (2009).

9.	J. Chang, L. F. Register, and S. K. Banerjee, Proc. Device Res. Conf., 31 (2012)

10.	Q. Zhang, G. Iannaccone, and G. Fiori, Electron Dev. Lett. 35, 129 (2014).

11.	K. -T. Lam, X. Cao, and J. Guo, Electron Dev. Lett. 34, 1331 (2013).

12.	H. Liu, A. T. Neal, Z. Zhu, Z. Luo, X. Xu, D. Tománek, and P. D. Ye, ACS Nano, 8, 4033 (2014).

13.	L. Li, Y. Yu, G. J. Ye, Q. Ge, X. Ou, H. Wu, D. Feng, X. H. Chen, and Y. Zhang, Nat. Nanotechnology, 9, 372 (2014).

14.	Y. Cai, G. Zhang, and Y. Zhang, Sci. Rep. 4, 6677 (2014).

15.	V. Tran, R. Soklaski, Y. Liang, and Li. Yang, Phys. Rev. B 89, 235319 (2014).

16.	S. Das, W. Zhang, M. Demarteau, A. Hoffmann, M. Dubey, and A. Roelofs, Nano Lett., 14, 5733 (2014).

17.	J. Qiao, X. Kong, Z. Hu, F. Yang, and W. Ji, Nat. Communications 5, 4475  (2014).

18.	X. Peng, Q. Wei, and A. Copple, Phys. Rev. B 90, 085402 (2014).

19.	T. Ozaki, and H. Kino, Phys. Rev. B 72, 045121 (2005).

20.	J. P. Perdew, and Y. Wang, Phys. Rev. B 45, 13244 (1992).

21.	A. Brown, and S. Rundqvist, Acta Cryst., 19, 684 (1965).

22.	X. Wang, A. M. Jones, K. L. Seyler, V. Tran, Y. Jia, H. Zhao, H. Wang, Li. Yang, X. Xu, and F. Xia, arXiv:1411.1695 (2014).

23.	H. Weng, T. Ozaki, and K. Terakura, Phys. Rev. B 79, 235118 (2009).

24.	H. Asahina, and A. Morita, J. Phys. C: Solid State Phys. 17, 1839 (1984).

25.	T. Usuki, M. Saito, M. Takatsu, C. R. Kiehl, and N. Yokoyama, Phys. Rev. B 52, 8244 (1995).

26.	Process Integration, Devices, and Structures (PIDS), *International Technology Roadmap for Semiconuctors* (ITRS) [Online]. Available: Http://www.itrs.net/

27.	J. Chang, L. F. Register, and S. K. Banerjee, J. Appl. Phys. 115, 084506 (2014).

28. Y. Yoon and S. Salahuddin, Appl. Phys. Lett. 101, 263501 (2012).S